\pdfoutput=1
\documentclass{iopart}

\usepackage[pdftex]{graphicx}
\usepackage{graphicx}

\usepackage{amscd,amssymb}
\usepackage{epstopdf}
\usepackage{latexsym}

\newcommand{\la}[1]{\stackrel{\leftarrow}{#1}}
\newcommand{\ra}[1]{\stackrel{\rightarrow}{#1}}
\newcommand{\ta}[1]{\stackrel{\leftrightarrow}{#1}}

\begin{document}

\title[Stability of Maximum likelihood based clustering methods]{Stability of Maximum likelihood based clustering methods: exploring the backbone of classifications}

\author{Muhittin Mungan$^{1,2}$ and Jos\'e J. Ramasco$^3$}

\address{$^1$Department of Physics, Bo\u gazi\c ci University, 34342 Bebek, Istanbul, Turkey\\
$^2$The Feza G\"ursey Institute, P.O.B. 6, \c Cengelk\"oy, 34680 Istanbul, Turkey\\$^3$Complex Networks Lagrange Laboratory (CNLL), ISI Foundation, 10133 Turin, Italy\\E-mail: mmungan@boun.edu.tr and jramasco@isi.it}

\date{Received:  / Accepted:  / Published }

\begin{abstract}
Components of complex systems are often classified according to the way they interact with each other. In graph theory such groups are known as clusters or communities. Many different techniques have been recently proposed to detect them, some of which involve inference methods using either Bayesian or Maximum Likelihood approaches. 
In this article, we study a statistical model designed for detecting clusters based on connection 
similarity. The basic assumption of the model is that the graph was generated by a certain grouping of the nodes and an Expectation Maximization algorithm is employed to infer that grouping. We show that the method admits further development to yield a stability analysis of the groupings that quantifies the extent to which each node influences its neighbors group membership. Our approach naturally allows for the identification of the key elements responsible for the grouping and their resilience to changes in the network. Given the generality of the assumptions underlying the statistical model, such nodes are likely to play special roles in the original system. We illustrate this point by analyzing several empirical networks for which further information about the properties of the nodes is available. The search and identification of stabilizing nodes constitutes thus a novel technique to characterize the relevance of nodes in complex networks.
\end{abstract}


\section{Introduction}

Networks are useful tools to characterize complex systems~\cite{newman-leicht,us,rives02,newman02,doreian,lorrain71,babu04,neo00}.
The system components are represented as nodes and their mutual 
interactions as edges. Finding structures in such networks is therefore of great relevance for understanding the mechanisms that underlie the system evolution. This explains the increasing interest in the topic, particularly in the detection of communities~ \cite{newman-leicht,us,newman02,radicchi,santo-pnas,newman-girvan,newman04,newman06,arenas,santo09,clauset}. 
Communities are groups of nodes with a high level of group inter-connection \cite{newman02}. They can be seen as relative isolated subgraphs with few contacts with the rest of the network. Communities have an obvious significance for social networks where they correspond to groups of close friends or well-established teams of collaborators \cite{lorrain71}. However, they are also important for characterizing other real-world networks such as those coming from biology~\cite{rives02,newman02,babu04} or from technology and transport~\cite{flake02,guimera-air,capocci}. 
Communities are not the only meaningful structures in networks: in Ecology, Computer and Social Sciences structurally equivalent nodes have been also considered~\cite{doreian,lorrain71,neo00}. These nodes are characterized by similar connectivity patterns and are expected to play similar roles within the system.  

There has been a long tradition of applying Bayesian and Maximum Likelihood methods to structure detection in networks~\cite{holland76,wang87,Snijders,hastings06,Airoldi,vazquez08,vazquez08b,zanghi08,hofman08}. These methods have the advantage that, depending on the statistical model used, they can be very general detecting both communities and structural equivalent set of nodes. The drawback, shared with many other methods, is that structure detection usually implies computational expensive exploration of the solutions maximizing the posterior probability or the likelihood. Recently, a maximum likelihood method that considers node clustering as missing information and deals with it using an Expectation Maximization (EM) approach has been introduced by Newman and Leicht~\cite{newman-leicht,us}. This method is computationally less costly to implement and we will denote it by the acronym NL-EM from now on. NL-EM is able to identify network structure relying on three basic assumptions: ({\it i}) the actual connectivity of the network is related to a coherent yet {\it a priori} unknown grouping of the nodes, ({\it ii}) the presence or absence of a link is independent from the other links of the network and ({\it iii}) the groups are tell-tales of processes that gave rise to the graph. No extra information is assumed except for the network itself and the number of groups. Under these assumptions, the method infers the classification of nodes that most likely generated the graph detecting communities and also structurally equivalent sets of nodes \cite{us}. Here we will show that due to the simple structure of the NL-EM likelihood, its classifications are based on a subset of nodes which turn out to be responsible for establishing the group memberships of their neighbors. We are able to rank the nodes according to the amount of group-allocation information they transmit to their neighbors and thereby identify those that are essential for establishing each group. These nodes, which we will refer to as stabilizers, constitute the backbone of the classification: the classification would not be viable without them and conversely, stabilizers turn out to emerge as a result of their distinct connection patterns on the given graph. Given the generality of the NL-EM underlying assumptions and that the resulting classifications can be validated by comparison with other clustering methods, we suggest that the stabilizers have an important inherent value for understanding the processes that generated the given network. Such an expectation is supported by our results on empirical graphs for which additional information regarding the nodes intrinsic properties is available. We will also briefly discuss the extension of this concept to other inference methods such as Bayesian clustering techniques~\cite{Snijders,Airoldi}.

\section{NL-EM clustering method}

We begin with a quick summary of NL-EM as applied to graphs. Labeling the nodes as $i = 1, \cdots, N$, the  variables are: $\pi_r$, the probability that a randomly selected node is in group $r$, $\theta_{rj}$, the probability that an edge leaving group $r$ connects to node $j$, and $q_{ir}$, the probability that node $i$ belongs to group $r$. The method is a mixture model where an edge between nodes $i$ and $j$ (expressed as $i \leftrightarrow j$) given 
the groups of $i$ and $j$ ($g_i$ and $g_j$) is observed with probability 
\begin{equation}
{\rm Pr}(i \leftrightarrow j \vert g_i, g_j) = \theta_{g_ij}\theta_{g_ji}. 
\label{eqn:thetacon}
\end{equation}
The edges are considered as independent so the probability that a given grouping realizes an observed network $\mathcal{G}$ can  be written as
\begin{equation}
{\rm Pr}(\mathcal{G} \vert \theta,\pi, \{g_i\}) = \prod_{i}  \pi_{g_i} \left [ \prod_{j \in \nu_i} 
\theta_{g_i j} \right ], 
\label{eqn:ProbA}
\end{equation}
where  $\nu_i$ is the set formed by the neighbors of node $i$. 

The group assignment captured by the terms $q_{ir}$  is treated as missing information.  The Expectation step of EM can thus be implemented as an average over the log-likelihood
\begin{equation}
\bar{\mathcal{L}}(\pi,\theta) = \sum_{ir} q_{ir} \left [ \ln \pi_{r} + \sum_{j \in \nu_i} \ln \theta_{rj} \right ] .
\label{loglik}
\end{equation}
The maximization of $\bar{\mathcal{L}}(\pi,\theta)$ is subject to the normalization
constraints, 
\begin{equation}
\sum_j \theta_{rj} = 1 \mbox{  and   } \,\, \sum_r \pi_r = 1 ,
\end{equation}
and leads to 
\begin{equation}
\begin{array}{rcl}
\theta_{ri} &=& \frac{\sum_{j \in \nu_i}   q_{jr}}{\sum_{j \in \nu_i} k_j q_{jr}} \\
\, & \, &\, \\
\pi_r &=& \frac{1}{N} \sum_i q_{ir} ,
\end{array}
\label{eqn:EM}
\end{equation}
where $k_j$ is the degree of node $j$. The group assignment probabilities $q$ are determined {\it a posteriori} from 
\begin{equation}
q_{ir} = \frac{ {\rm Pr}(\mathcal{G}, g_i = r \vert \theta,\pi)}{{\rm Pr}(\mathcal{G} \vert \theta,\pi )} ,
\end{equation}
as
\begin{equation}
q_{ir} = \frac{\pi_r \prod_{j \in \nu_i} \theta_{rj}}{\sum_s \pi_s \prod_{j \in
\nu_i} \theta_{sj}}.
\label{eqn:qir}
\end{equation}
The maximization of $\bar{\mathcal{L}}$ can be carried out with different techniques. In order to 
account for the possible existence of a rough likelihood landscape with many local extrema, 
we employed an algorithm that alternates between simulated annealing and direct greedy iteration of Eqs.~(\ref{eqn:EM}) and (\ref{eqn:qir}). 

\section{Stability analysis and stabilizers}   
\label{RelStab} 

The group membership of the nodes is encoded by the probabilities $q$. It is thus natural to ask for the conditions on a node $i$ and its neighbors to have $i$ crisply classified into a single group $r$ so that $q_{is} = \delta_{rs}$. The answer highlights the role of the neighbors in establishing a node's membership credentials. Looking at the expression for $q_{ir}$, Eq.~(\ref{eqn:qir}), 
\begin{equation}
q_{ir} \sim  \prod_{j \in \nu_i} \theta_{rj},
\end{equation}
where the non-zero prefactors whose sole role is to ensure proper 
normalization have been suppressed, 
one finds that for each group $s \ne r$ there must be at least one neighbor $j$ of $i$ whose probability $\theta_{sj}$ is zero. However, as seen from Eq.~(\ref{eqn:EM}), whether $\theta_{sj}$ is zero or not for some group $s$ depends in turn on the group memberships of the 
neighbors of $j$. Hence having a node crisply classified as belonging to a group sets strong constraints on its neighbors and their respective neighborhoods. These constraints propagate throughout the network during the NL-EM iteration until a final configuration for $\theta$ and $q$ is established. 
In this sense, a node $j$ is passing information about group 
membership to its neighborhood through the probabilities $\theta_{sj}$. This information is negative, of the form "you do not belong to group $X$" when  $\theta_{Xj}$ is zero and we say that node $j$ stabilizes its neighbors against membership in group $X$. It is worth noting the parallels of this mechanism with message passing algorithms~\cite{mackay}. In a classification into 
$\mathcal{N}_C$ groups each crisply classified node $i$ must be stabilized against $\mathcal{N}_C - 1$ groups. Thus one can regard the number of groups a node $j$ 
stabilizes against as a measure of the amount of information $I_j$ that $j$ passes to its neighbors. If $I_j = \mathcal{N}_C - 1$, node $j$ can stabilize its adjacent nodes alone providing thus complete information about their group membership. On the other hand,  when $I_j < \mathcal{N}_C - 1$, $j$ provides only partial information. 
%
%
The crisp classification of a neighbor $i$  requires then the combined action of other adjacent nodes in order to attain full group membership information. We denote as {\it stabilizers} of $i$ the union set of neighbors that alone or in combined action pass essential information to $i$ establishing its membership in a single group (a more precise definition will be given below).  The above analysis implies that any crisply classified node must be stabilized by one or more stabilizers. 
Therefore, if the assumptions of the statistical model are justified and the resulting node classification is meaningful, the identification of the corresponding stabilizers may offer useful additional information. 

Based on their classification and the information passed, four types of nodes can be distinguished: nodes can be strong or weak depending on whether they are crisply classified into a single group or not, and 
they can be stabilizers or not, depending on whether they pass essential information for establishing an adjacent  node's group membership.
If we consider a node $i$ and denote by  
$\bar{\sigma}_i = \{r \vert \theta_{ri} = 0\}$, the set of groups that  $i$ does not connect to,  and by $\bar{c}_i = \{r \vert q_{ir} = 0\}$, the set of groups that $i$ does not belong to, the NL-EM equations (\ref{eqn:EM}) and(\ref{eqn:qir}) relate these sets as follows:
\begin{equation}
\bigcup_{j \in \nu_i} \; \bar{\sigma}_j = \bar{c}_i \;\;\;\;\; \mbox{and} \;\;\;\;\;
\bigcap_{j \in \nu_i} \; \bar{c}_j = \bar{\sigma}_i, 
\label{eqn:sets}
\end{equation}
forming a set of consistency relations with a simple meaning: a node cannot belong to a group to which its neighbors do not connect, and the common set of groups to which a node's neighbors do not belong must correspond to the groups that it does not connect to. If we require in particular that a node $i$ is strong, {\it i.e.} it is crisply classified as belonging to a particular group $A$, then $\bar{c}_i = \mathcal{C} \setminus \{A\}$ \cite{NOTE}.

\begin{figure*}[h]
\begin{center}
\includegraphics[width=6cm]{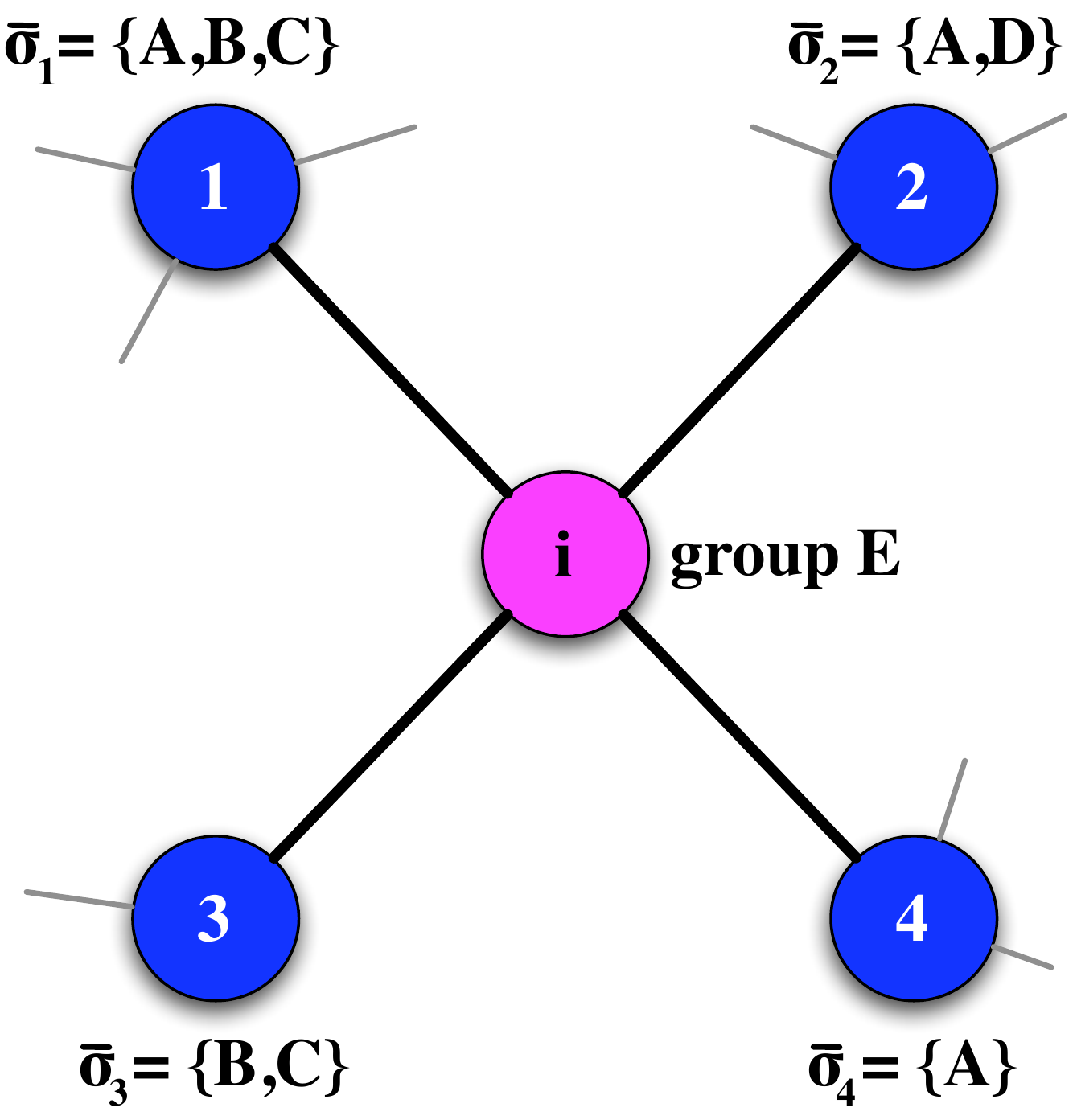}
\caption{Stabilization of a strong node $i$. The groups are $\mathcal{C} = \{A,B,C,D,E\}$ and $i$ is crisply classified as $E$. The four adjacent nodes are shown along with the set of classes $\bar{\sigma}$ to 
which they have no connections to. In order for $i$ to be classified as $E$ there must be a subset of adjacent nodes such that the union of their corresponding $\bar{\sigma}$ is $\mathcal{C}-\{E\}$, Eq.~(\ref{eqn:sets}). All four adjacent nodes must stabilize $i$, as otherwise $i$ would not be a strong node. However, the sets $\{1,2\}$ or $\{2,3\}$ each suffice to stabilize node $i$. The node $4$ is redundant, in the sense that any stabilization of $i$ involving node $4$ remains a stabilization of $i$  when $4$ is removed. A precise definition of stabilizers is given in the text.}
\label{fig:example}
\end{center}
\end{figure*} 

Given the sets $\bar{\sigma}_j$ 
associated with the neighbors $j$ of a strong node $i$, not all adjacent nodes need to contribute
to its full stabilization. Likewise, node $i$ can be stabilized by different combinations of its neighbors' 
sets $\bar{\sigma}_j$. This is best illustrated by an example shown in Fig.~\ref{fig:example}. 
Suppose that the groups are $\mathcal{C} = \{A,B,C,D,E\}$
and let us assume that node $i$ is crisply classified as $E$. Let $i$ have four neighbors with 
corresponding sets $\bar{\sigma}_1 = \{A,B,C\}$, $\bar{\sigma}_2 = \{A,D\}$, $\bar{\sigma}_3 = \{B, C\}$ and $\bar{\sigma}_4 = \{A\}$. 
It is clear that all four nodes together must 
stabilize $i$, as otherwise $i$ would not be a strong node. However, the sets of neighbors $\{1,2\}$ or $\{2,3\}$ each suffice to stabilize node $i$. The node $4$ is redundant, since it does not contribute a new class against which $2$ or $3$ are not already stabilizing. In other words, if the set $\{ 2,3,4 \}$ is considered, node $4$ can be removed without altering the stabilization of $i$. The same 
is not true for the nodes $2$ and $3$. 
The notion of stabilization sets and stabilizer nodes can be defined as follows: A subset of nodes adjacent to $i$ is a stabilization set of $i$, if the removal of any one of the nodes from the set causes $i$ not to be stabilized by that set anymore. A node $j$ is a stabilizer if it is member of at least one stabilization set. The definition of stabilizer involves thus a stabilization relation with at least one of the node neighbors. In the above example, $1, 2$ and $3$ are the only stabilizers of $i$. 
Non-stabilizer nodes can be removed without affecting stabilization, while whenever a stabilizer is removed the number of ways in which a given node is stabilized decreases. In the example of Fig.~\ref{fig:example}, the removal of node $2$ would cause complete loss of stabilization of $i$, while removal of $3$ or $1$ would leave 
$i$ with only a single stabilization. 
It can be shown that the removal of a stabilizer 
will never turn a previously non-stabilizer node into a stabilizer, but it might turn some  
stabilizers into non-stabilizers. 
Note that in a sense stabilizer $2$ is more important than $1$ or $3$, since it is part of every stabilization of $i$ and its removal will thus render $i$ a weak node. 
In fact, one could attach a strength to each stabilizer by keeping track of the number of stabilizations in which it is 
involved, but, for sake of simplicity, we will not pursue this here.

Given an NL-EM classification with strong nodes, we can immediately identify the stabilizers that are responsible for the crisp classifications. Details on how to implement the identification of stabilizers 
are provided in Appendix A.2. The relation $i$ stabilizes $j$ induces a directed subgraph on the original 
network and we will refer to this as the stabilizer subgraph. The relation between two stabilizer nodes is not necessarily of mutual stabilization:  a necessary condition for adjacent strong nodes $i$ and $j$ to mutually stabilize each other is that both $\bar{\sigma}_i \cap c_j$ and $\bar{\sigma}_j \cap c_i$ are empty. 
The connections among strong stabilizers capture the relations between groups in the graph. In that sense one can regard the stabilizers as exemplary members of the groups. In the undirected graphs of Figs.~\ref{fig:random} - 
\ref{fig:adj} the stabilizer subgraph has been superposed. The extension of these concepts to NL-EM classifications in directed graphs is similar, details are given in Appendix B. 

The case of NL-EM classifications into two groups is particularly simple. Denoting the groups as $A$ and $\bar{A}$, a crisply-classified (strong) node belongs to either $A$ or $\bar{A}$ and a strong node of a given group has to be stabilized against the complementary group. All nodes with non-empty $\bar{\sigma}$ are therefore stabilizers, and if more than one is present all are equivalent, each stabilizing a given node independently from the other stabilizers. Moreover, the strong stabilizers are nodes that are stabilized themselves by some of their neighbors which necessarily are also stabilizers. The conditions of Eq.~(\ref{eqn:sets}) permit only two possible configurations of the stabilizer subgraphs. Either strong stabilizers of group $A$ connect to strong stabilizers of their own group, or stabilizers of group $A$ connect to those of the complementary group $\bar{A}$. In the former case we get a disjoint community like partition ({\it cf. } Fig.~\ref{fig:sen}) of the stabilizer graph, whereas in the latter case  we obtain a bipartite partition ({\it cf. } Fig.~\ref{fig:adj}).

Furthermore, the NL-EM classification into two groups reveals a simple but meaningful hierarchical structure in the way the different type of nodes in the classification relate. Strong (non-stabilizer) nodes are nodes for 
which $\bar{\sigma} = \emptyset$, so these nodes connect to nodes of both groups (weak 
or strong), however in order for them to be strongly classified as in one group, let us say, $A$ ($\bar{A}$) they 
can only connect to those stabilizer nodes with the compatible stabilizer classes 
$\bar{\sigma} = \{ \bar{A}\}$ ($\bar{\sigma} = \{ A \}$). In turn, the neighborhood of 
strong stabilizer nodes with $\bar{\sigma} = \{ A\}$ or $\bar{\sigma} = \{ \bar{A}\}$ can 
consist only of nodes strongly classified as $\bar{A}$ or $A$, respectively. The weak 
stabilizer nodes are by definition nodes for which $\bar{c} = \emptyset$, but for which
$\bar{\sigma} = \{ A\}$  or $\bar{\sigma} = \{ \bar{A}\}$. Thus weak stabilizer nodes 
cannot connect to strong stabilizer nodes, but they can stabilize strong (non-stabilizer) 
nodes. Finally, the weak nodes that are neither strong nor stabilizing can connect to 
strong non-stabilizing nodes and other weak nodes. In this way the connection rules for 
the strong stabilizers, weak stabilizers, strong nodes, and weak nodes set up a hierarchy 
of nodes at the core of which are the strong stabilizers.

\section{Stabilizers in a benchmark}

\begin{figure*}[t]
\begin{center}
\includegraphics[width=\textwidth]{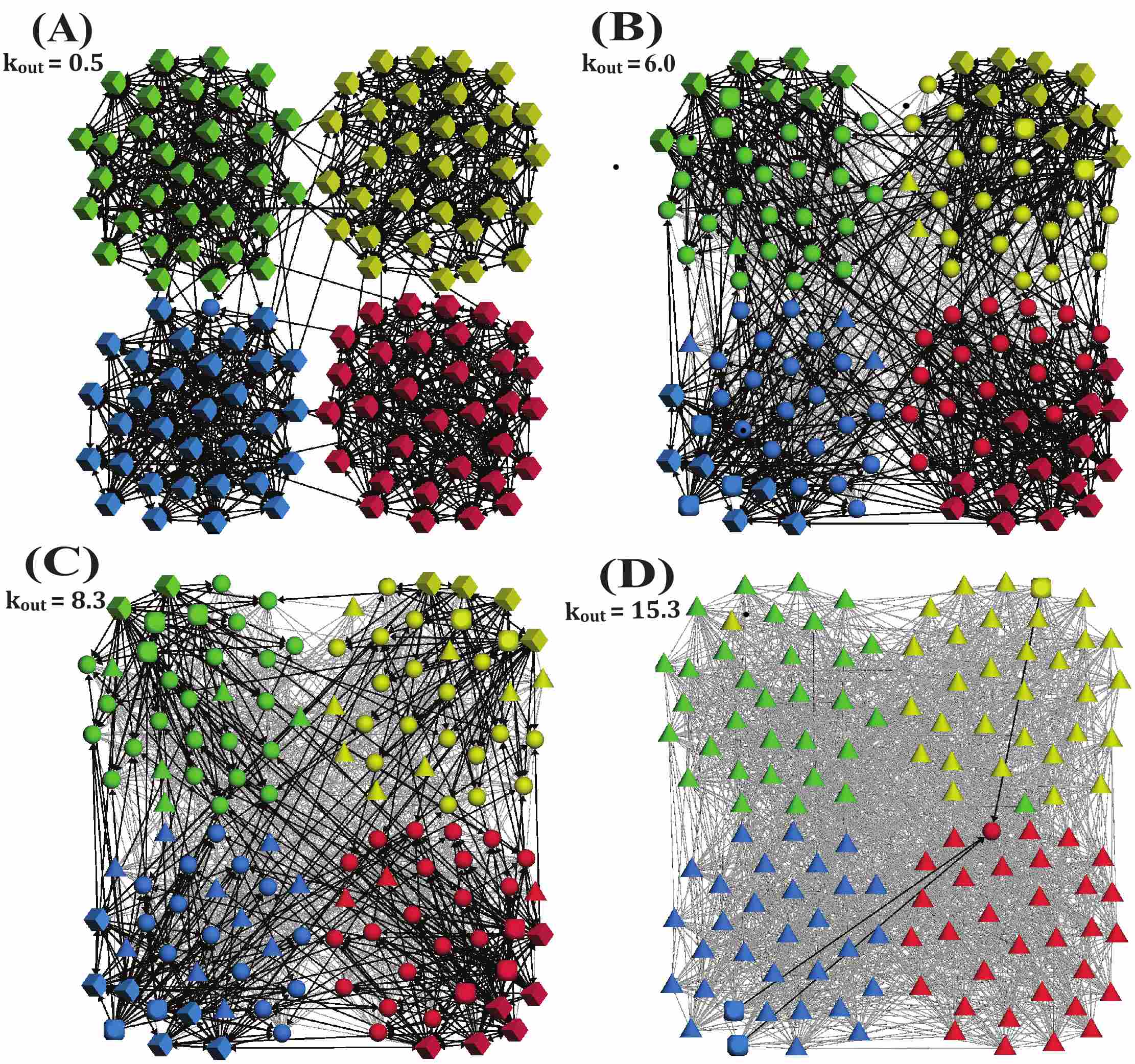}
\caption{Benchmark with four communities and increasing intra-community connections, $k_{out}$. (A), (B), (C) and (D): four instances of the graph classification: strong stabilizers, weak stabilizers, strong nodes and weak nodes are shown as rhomboids, cubes, spheres and cones, respectively. Nodes have been arranged by communities while their color depicts the group to which the NL-EM algorithm assigns them with highest probability. 
The directed (dark) arcs show the information flow, as captured by the stabilization relation. }
\label{fig:random}
\end{center}
\end{figure*}

As we observed in the previous section, a node can be stabilized by its neighbors in multiple ways. This redundancy renders classifications robust against disorder introduced by the addition or removal of edges up to a certain point. To illustrate this we consider 
a benchmark with four communities~\cite{newman-girvan}. The initial network is generated with four disjoint groups of $32$ nodes each, with the nodes having on average $\langle k_{in} \rangle = 16$ in-group links. These groups correspond to the four clusters of Fig.~\ref{fig:random}(A)-(D).  Random links connecting different groups are added to the basic configuration and the number of stabilizers are tracked as a function of the average number of out-group links $k_{out}$. Fig.~\ref{fig:random} shows the stabilizers obtained from an NL-EM classification into  $\mathcal{N}_C = 4$ groups at disorder level $k_{out} = 0.5, 6.0, 8.3$ and $15.3$. 

\begin{figure*}[t]
\begin{center}
\includegraphics[width=\textwidth]{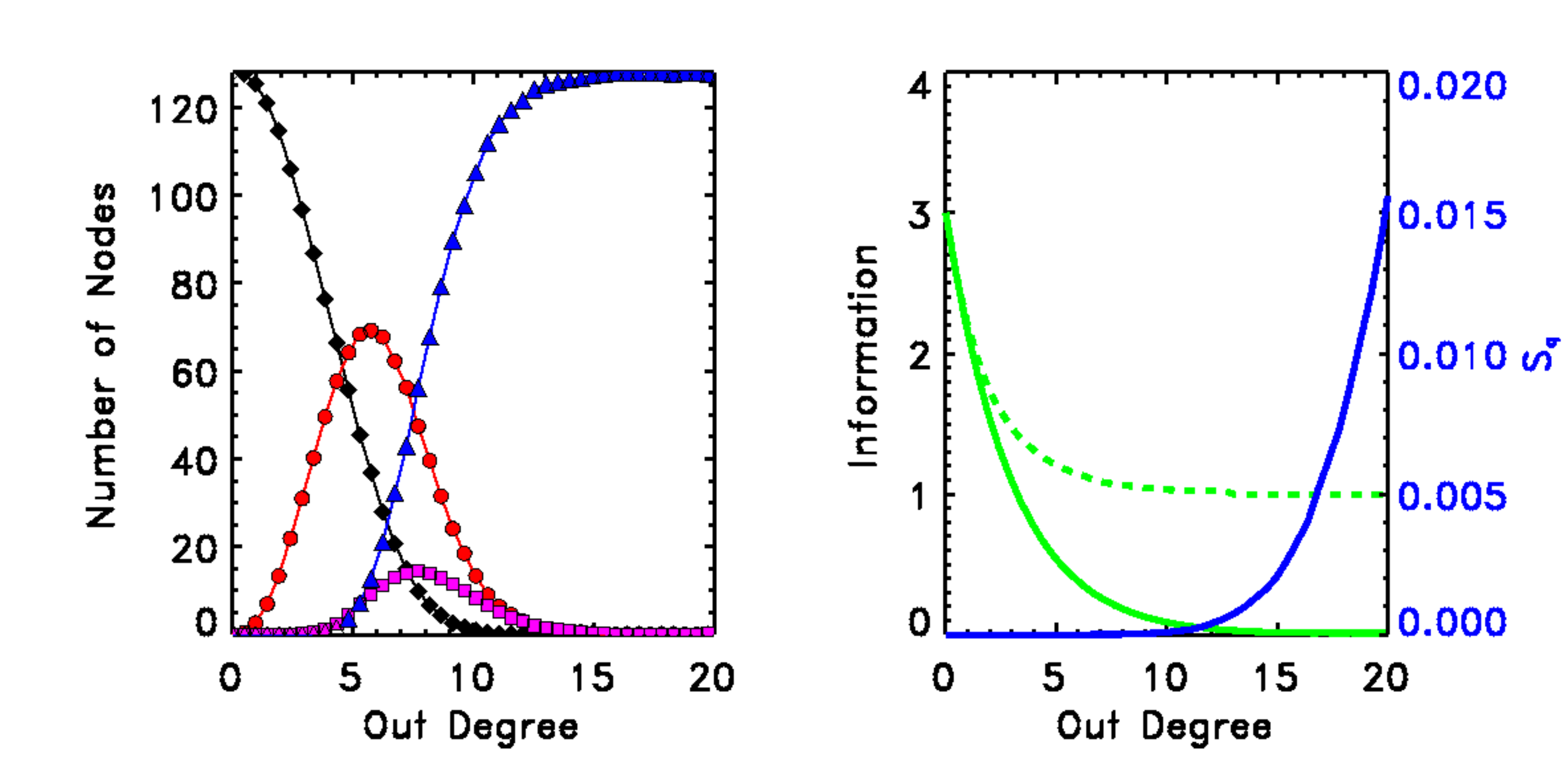}
\caption{Benchmark with four communities and increasing intra-community connections continued. (Left) Number of nodes vs. disorder: strong (diamonds) and weak (boxes) stabilizers, strongly (circles) and weakly classified nodes (triangles). (Right) average information $\langle I \rangle$ passed by all nodes (continuous green curve) or stabilizers only (green dashed curve, for which $I \ge 1$ by definition). The blue curve shows the entropy of the classification, as 
defined in Eq.~(\ref{eqn:Sq}). The values for each data point in the plots have been obtained from averaging over 100 realizations of the random process of edge additions.}
\label{fig:random_class}
\end{center}
\end{figure*}

When $k_{out} = 0$ we find a crisp classification where all nodes are strong stabilizers, 
meaning that all nodes stabilize and are being stabilized. Furthermore, all of them provide complete stabilization information, $I = 3$, with a single stabilizer sufficing to crisply classify a neighbor. Since $\langle k_{in} \rangle = 16$, there is on average 16-fold redundancy in the stabilization of each node. 
As random connections are added to the network, the four clusters become 
connected with each other. Some of the stabilizers start to stabilize 
against fewer classes, giving rise to a decrease in the average $I$. 
In the right panel of Fig.~\ref{fig:random_class}, we have plotted how the average stabilization information decays when $k_{out}$ increases. In order for nodes with $I <3$ to be stabilizers they have to act in combined action with other nodes, as in the example of Fig.~\ref{fig:example}.
Thus an increase of  the level of disorder $k_{out}$ causes both a reduction in the redundancy of 
stabilizations of strong nodes and a shift towards stabilizations by combined action of more than one stabilizer. The increase in disorder eventually leads to a loss of strong nodes, implying that the classification deteriorates. In order to assess the quality of classifications, we use the  entropy $S_q$, as defined in \cite{us}
\begin{equation}
S_q = - \frac{1}{N} \sum_{ir} q_{ir} \ln q_{ir}.
\label{eqn:Sq}
\end{equation}
The entropy $S_q$ measures the crispness of a classification. When $S_q = 0$, all the nodes are strong, while $S_q = \ln(\mathcal{N}_C)$ corresponds to case where the classification of the nodes is 
maximally uncertain. The right panel of Fig.~\ref{fig:random_class} displays $S_q$
as a function of $k_{out}$, showing that the crispness of the classification is lost for 
large $k_{out}$.

The increase in entropy is closely related to what happens to the different nodes in the classification as edges are added, particularly to the stabilizers. The variation of the number of the different type of nodes with $k_{out}$ is shown in the left panel of Fig.~\ref{fig:random_class}. As the addition of new edges progresses, some nodes cease to be strong stabilizers. 
When a node is not a strong stabilizer anymore, it can still remain strong as long as there are other nodes stabilizing it in its neighborhood. As can be seen in the left panel of Fig.~\ref{fig:random_class}, this is 
what is happening up to $k_{out} \lesssim 4$: The number of strong 
stabilizers decreases while the number of strong nodes rises accordingly. Therefore, initially the effect of adding edges is to convert strong stabilizers into 
strong nodes. Most of the nodes remain strong (stabilizer or not), and
the classification is essentially crisp with an entropy $S_q \approx 0$. 
With the further addition of edges, the number of strong nodes starts to decrease 
as a result of the loss of stabilization, giving rise to the appearance of 
weak stabilizing and non-stabilizing nodes at $k_{out} \gtrsim 4$. Continuing to $k_{out} \approx 10$, the entropy of the classification remains very low because there still is a 
sizable number of strong nodes supported by a few weak and strong stabilizers (see panels B and C in Fig.~\ref{fig:random}). As further edges are added, the number of 
weak stabilizers starts to decrease as well, and eventually most of the nodes are
weak and non-stabilizing, accounting for the quick rise in 
the classification entropy $S_q$ starting around $k_{out} \approx 10$.

\section{Real-world networks}

We focus now on some empirical examples to show the special role that the stabilizers play in a classification and the type of information that they convey while also highlighting the versatility of our analysis. As explained, classifications into two groups are particularly simple and in this case the stabilizers can be easily identified once a solution of the NL-EM clustering is given. This simplicity makes them good candidates to illustrate the properties of the stabilizers. We present first two examples of this type that show the role of the stabilizers and the relations between them. We then turn to a directed network with a classification into $4$ groups in order to illustrate a more general situation. 

\begin{figure}
\begin{center}
\includegraphics[width=\textwidth]{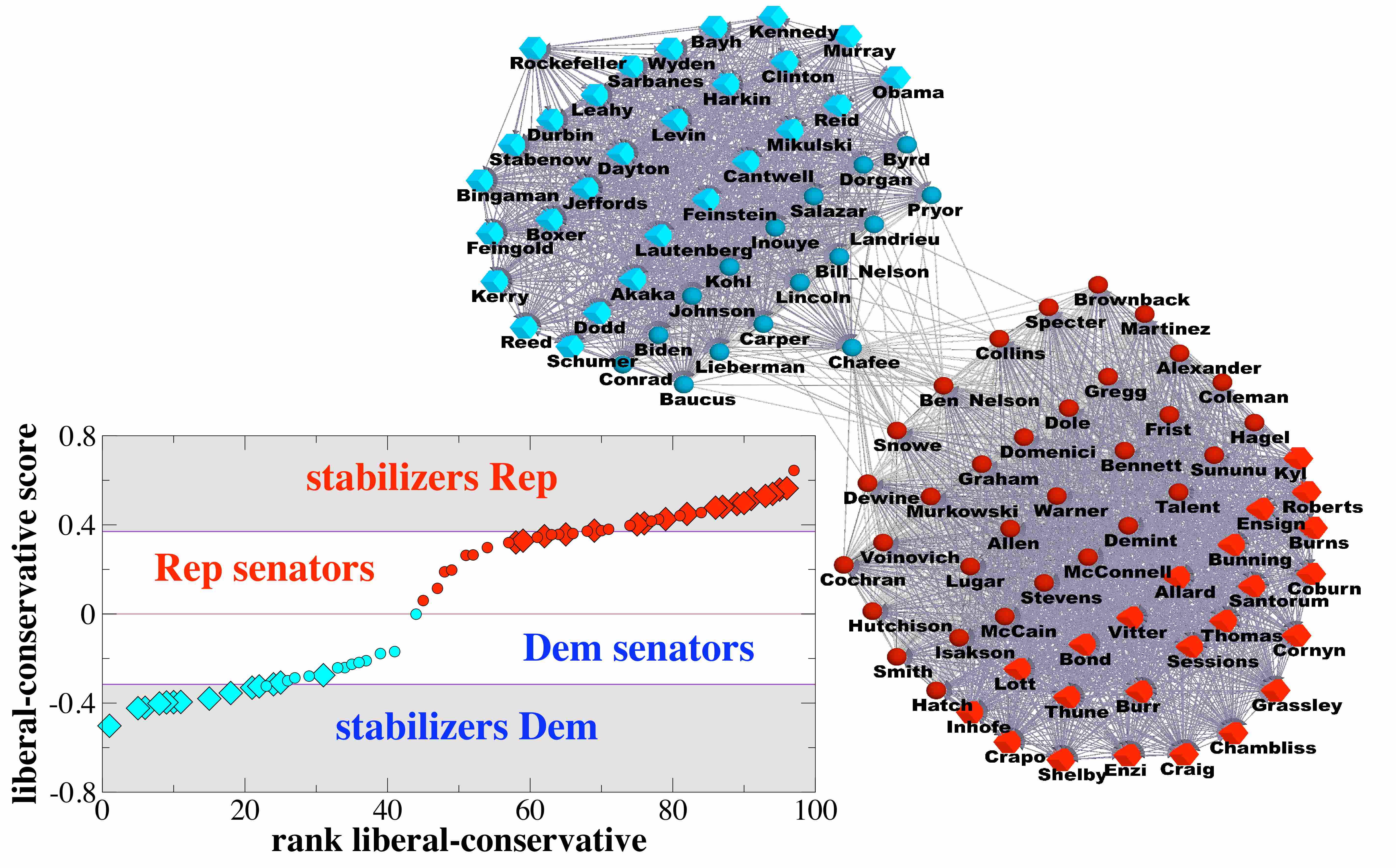}
\caption{Political affinity network between congressmen of the 109th US Senate. Right: The node shapes are as in Fig.~1: rhomboids are strong stabilizers and spheres well classified nodes not passing essential group information. The colors convey the NL-EM classification which follows the partition into 
Democrats (blue) and Republicans (red) fields. Left: the senators are displayed as ranked according to their liberal-conservative score \cite{poole07,poole05}. The average values of the score in the different sub-groups are: Rep. stabilizers $0.45\pm 0.08$, Rep. strong nodes $0.33\pm0.12$, Dem. strong nodes $-0.22\pm0.08$ and Dem. stabilizers $-0.37\pm0.06$.}
\label{fig:sen}
\end{center}
\end{figure} 

The first example is a network built from the voting records of the $109$th US Senate \cite{data_sen}. The nodes represent senators that served the full two year term ($2005-2007$) during which $645$ issues were voted. Since our aim is to construct a network based on political affinity, we draw an edge between two senators if they voted in the same way at 
least once. The edges are weighted by the reciprocal of the number of co-voting senators minus one, a common practice for collaboration networks \cite{newman01}. In this way, an agreement in minority on an issue has a higher value than that in an unanimous vote, differentiating more clearly close political standings. Due to circumstantial quasi-unanimous votes, the network is initially close to fully connected. A threshold such that edges with lower weights
are removed can be introduced, and the resulting networks can be analyzed as the threshold increases. We have applied two-group NL-EM to these networks. Once the threshold is high enough, the clusters found follow well the divide between Democrats and Republicans. The instance in which about half of the senators, either Republicans or Democrats, are stabilizers is displayed in Figure~\ref{fig:sen}. 
Congress roll calls and their derived networks have been extensively studied in the literature \cite{poole07,poole05,porter05,porter07}. One of the most interesting results is that single votes of a representative can be understood with a low dimensional spatial model (DW-NOMINATE \cite{poole07,poole05}) 
in which a set of coordinates can be assigned to each congressman characterizing his/her political stand on the different issues. Since the 90's the number of dimensions required has been reduced in good approximation to only one that strongly correlates with the congressman's view on socio-economic questions (liberal vs. conservative) \cite{poole07,poole05}. In Fig~\ref{fig:sen}, we show the relation between being a stabilizer and the location in the liberal-to-conservative dimension. The stabilizers tend to be the most radical members of the Senate who are probably defining the overall position of their groups. This exercise can be repeated on networks obtained with different thresholds. It can be seen that as the threshold increases more and more nodes turn into stabilizers. Keeping track of the senators that become stabilizers at different thresholds allows for a refined exploration of the political spectrum. Note in particular that the above results have been obtained by simply looking at the co-voting relation and without considering the vote records in detail, {\it i.e., the actual issue put to vote}.

\begin{figure}
\begin{center}
\includegraphics[width=\textwidth]{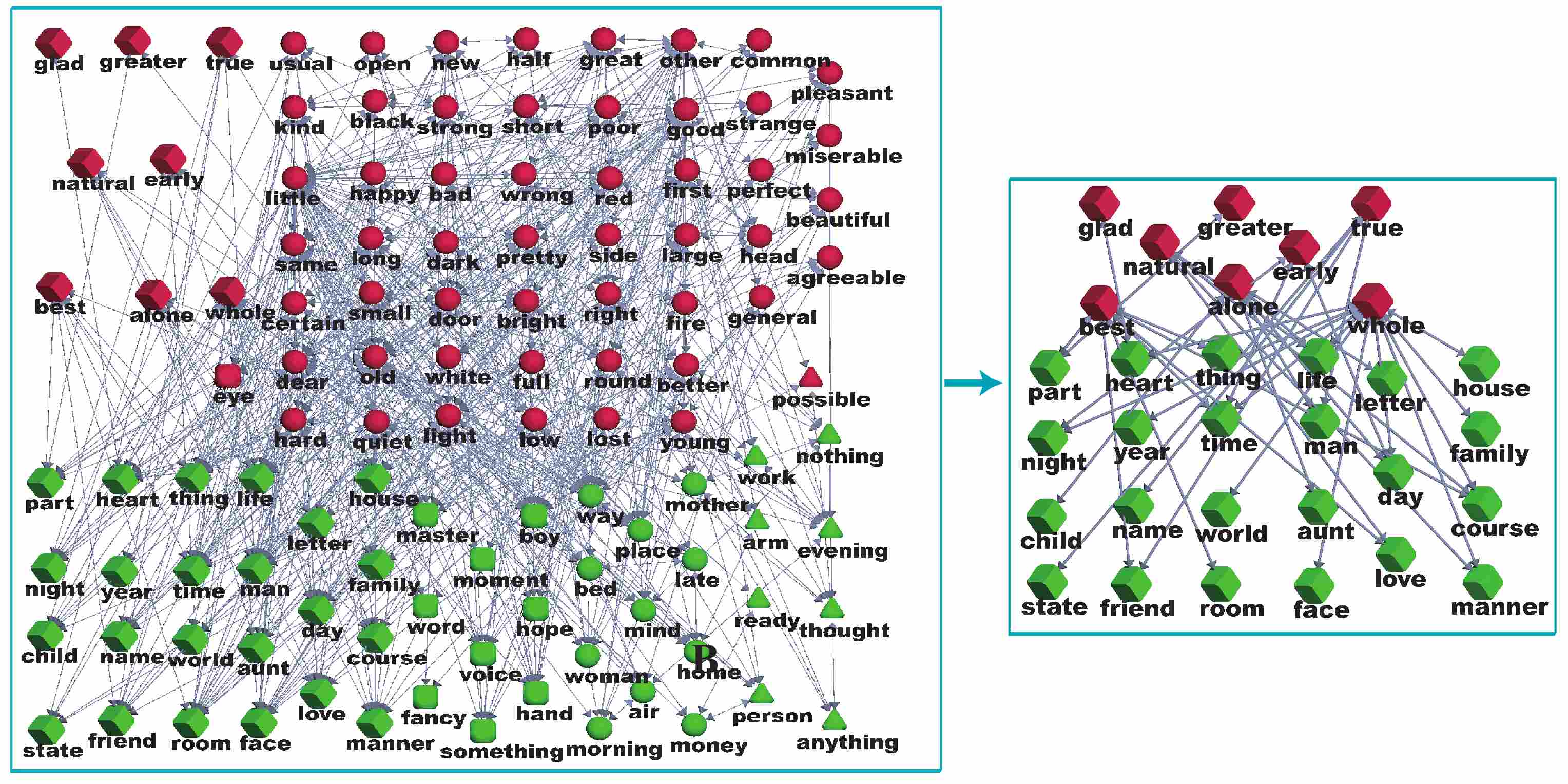}
\caption{Stabilizer analysis for the noun-adjective network in {\it David Copperfield} \cite{newman-adjnam}. The links represent words appearing in juxtaposition, while the superposed directed links indicate the stabilization relations. The NL-EM class assignment correlates strongly with the word being a noun (green) or an adjective (red). On the right, the subgraph formed by the strong stabilizers that exhibits a strict bipartite ordering.}
\label{fig:adj}
\end{center}
\end{figure} 

In our second example we show how by extracting the sub-graph of stabilizers we can obtain from its structure useful information about what features distinguish a stabilizer node and how the groups relate in a classification. We consider a semantic network in  which the nodes are adjectives and nouns occurring most frequently in Charles Dickens' novel {\it David Copperfield} \cite{newman-adjnam}. A relation between any two of these words is established if they occur in juxtaposition. In Fig.~\ref{fig:adj}, we have represented the network, the best NL-EM partition in two groups and identified the types of nodes. There turn out to be two sub-groups containing nouns or adjectives only that are strong stabilizers. These two sub-groups bear the responsibility for the classification of remaining words by association. Note that the only input to the NL-EM method is the network. We are not introducing any bias for the partition in adjectives and nouns. Most of the remaining words are well classified. The stabilizers, central to establishing the classification, are the words always occurring in strict combinations like {\it true friends}, never mixing with members of the same group and they form a bi-partite sub-graph of stabilizers as shown in the right panel of Fig.~\ref{fig:adj}. Conversely, nonstabilizing nodes are words appearing in mixed roles, such as the word {\it little} in the adjective-adjective-noun triplet {\it poor little mother}.

\begin{figure}
\begin{center}
\vspace{1.5cm}
\includegraphics[width=10cm]{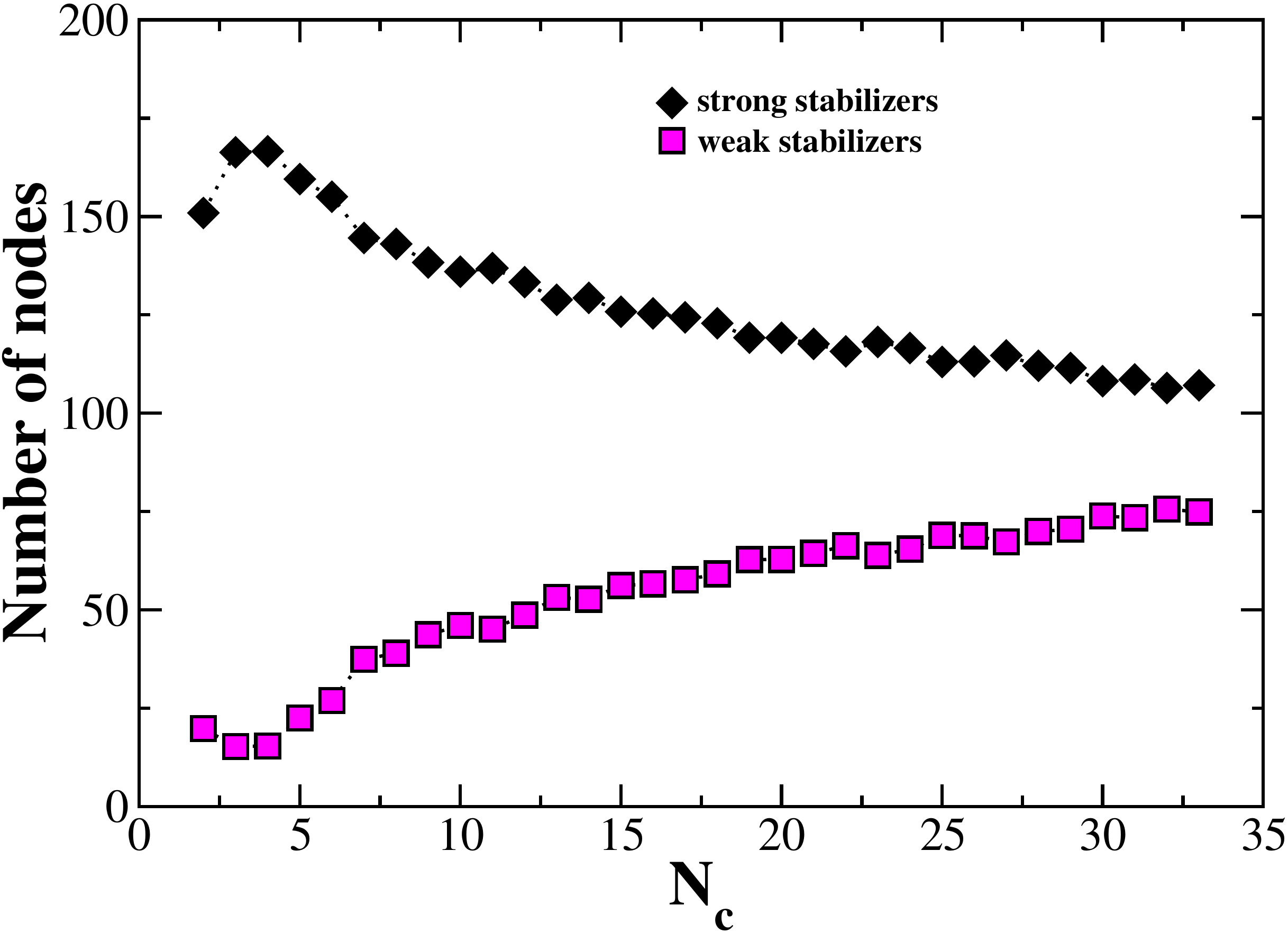}
\end{center}
\caption{Number of strong and weak stabilizers vs number of groups $\mathcal{N}_C$ as obtained from the NL-EM classification of the Little Rock Lake foodweb~\cite{neo91}. The maximum number of strong stabilizers occurs around 
$\mathcal{N}_C = 3, 4$. This number is close to the trophic level which is around $4$, suggesting that a classification 
into $4$ groups might capture the trophic levels.}
\label{fig:foodstab}
\end{figure}

\begin{figure}
\begin{center}
\vspace{-0.1cm}
\includegraphics[width=\textwidth]{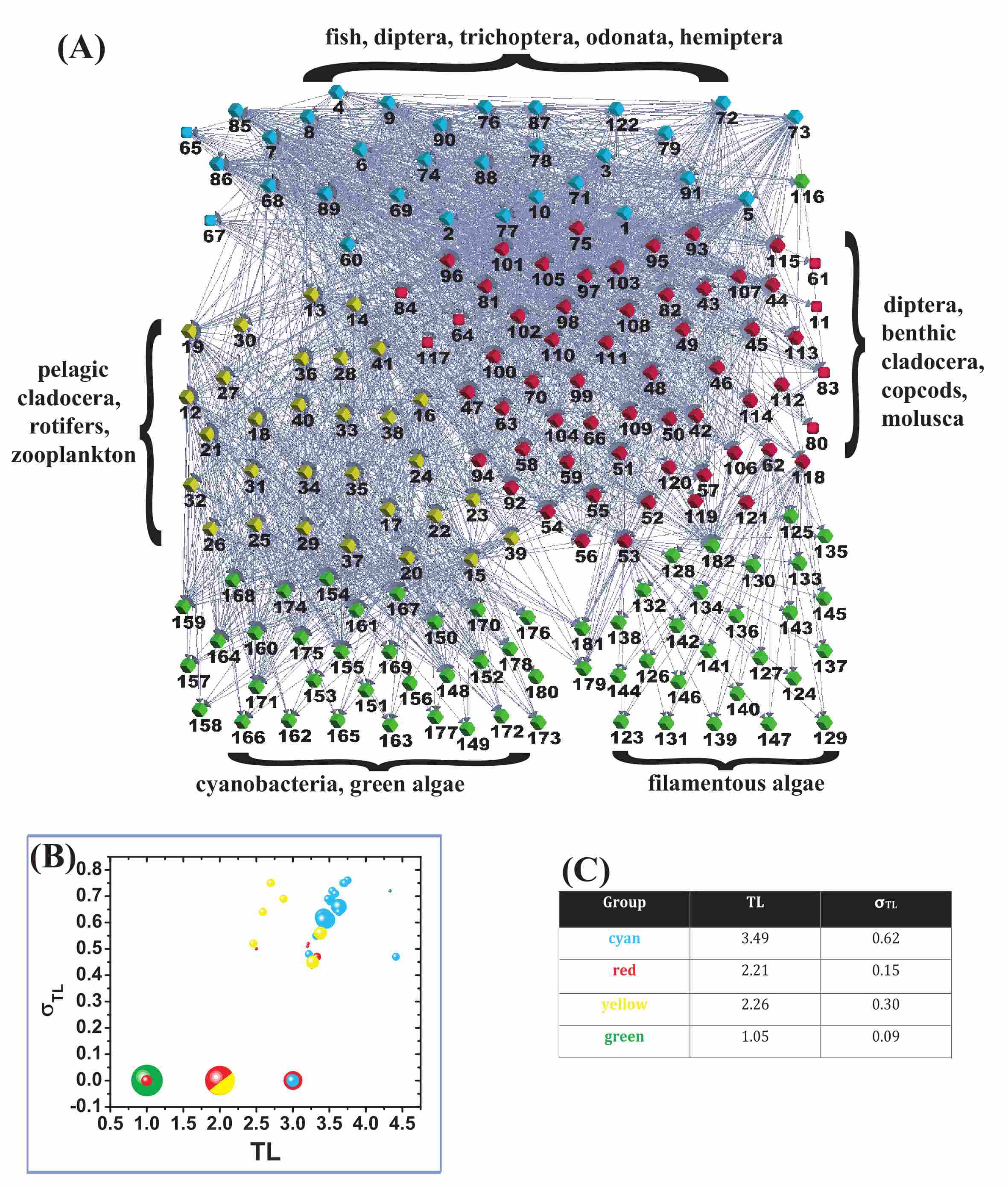}
\end{center}
\caption{(A) Stabilizers for the best $4$ group NL-EM classification of the Little Rock Lake foodweb~\cite{neo91}. Nodes are species and directed links correspond to predation relations. The node labeling follows \cite{neo91}. (B): fraction of species belonging to each group plotted against their prey-averaged trophic level ($TL$) and the standard deviation of $TL$ of their preys, as defined in \cite{neo04}. The radius of the spheres is proportional to the {\it log} of the percentile. Spheres with two colors include species of more than one group (each sphere or half-sphere is independent). (C): averages of $TL$ and $\sigma_{TL}$ over the species forming each group.}
\label{fig:food}
\end{figure}

Our final example, showing a more general case with $4$ groups, is the Little Rock food-web. The vertices of this network are species living in the aquatic environment of Little Rock Lake in Wisconsin~\cite{neo91}. Each directed link represents a predation relation pointing from predator to prey. The number of trophic levels is around four~\cite{neo04} and turns out to be the number of groups for which the NL-EM algorithm produces a partition with highest abundance of strong stabilizers, as shown in Fig.~\ref{fig:foodstab} where we have plotted the number of stabilizers of an NL-EM solution against the number of groups $\mathcal{N}_C$. A property of the four group classification 
depicted in Fig.~\ref{fig:food} is that it keeps basal species (green) in one group, top predators (cyan) in another, and assigns the rest to two different groups based on the prey they feed on at the basal level. The species that are not strong stabilizers, for instance nodes $11$, $61$ or $80$, could be related to a missing data problem. In the case of $61$ ({\it Hydroporus}) or $80$ ({\it Lepidoptera Pyralidae}), the species appear only as prey having no connection to lower levels. However, its consumers are not typically feeding on basal species, they are "cyan", and this results in an NL-EM classification that assigns them into the "red" group.

As seen in Fig.~\ref{fig:food}(A), most of the species of the network are strong stabilizers. Their abundance is a direct result of the highly structured organization of the foodweb: 
similar species have similar prey which, as our analysis shows, is also linked to their trophic levels (see Fig.~\ref{fig:food} B and C). Or more correctly, the consistent choice of species a predator does not prey on is what renders them stabilizers. The possibility of classifying species in low dimensional spaces depending on their trophic level and on the intervality of their prey distribution has been extensively discussed in the literature \cite{neo00,cattin04,stouffer06,allesina08}. 
Our stability analysis reveals an underling  
structure in the connectivity pattern of the foodweb, which is responsible for the success of these low dimensional models.

\section{Discussion and Conclusions}

The maximum likelihood function upon which the NL-EM inference method is based is rather generic and depends on the assumption that nodes with similar connections should be grouped together. Using this likelihood function 
we were able to show that a subset of nodes, the stabilizers, associated with a given grouping play a central role as they form the backbone of the classification which  
could not be attained without them. The mathematical basis behind the concept of stabilizers is rather intuitive and 
follows from the product form of the group assignment probabilities, $q_{ir}$, in Eq.~(\ref{eqn:qir}), 
which is in turn a direct consequence of the assumption that the edges are statistically independent (Eq.~(\ref{eqn:ProbA})). Such an assumption is common to a number of probabilistic clustering methods. We can rewrite 
Eq.~(\ref{eqn:qir}) as 
\begin{equation}
q_{ir} = \prod_{j \in \nu_i} \tilde{\theta}_{rj}, 
\end{equation}
where 
\begin{equation}
\tilde{\theta}_{rj} = \theta_{rj} \left [ \frac{\pi_r}{\sum_s \pi_s \prod_{j \in \nu_i} \theta_{sj}} \right ]^{\frac{1}{k_j}}, 
\end{equation}
so that the prefactors  are equally absorbed into $\tilde{\theta}_{rj}$. Note that $q_{ir}$ is in the interval $[0,1]$. Written in the above form it is clear that very small values of $q_{ir}$ must arise from very small values of $\tilde{\theta}_{rj}$ dominating the product. Likewise, we see from  
 ${\rm d}\ln q_{ir}/{\rm d}\tilde{\theta}_{rj} = 1/\tilde{\theta}_{rj}$ that changes in these factors 
will have the greatest effect on the value of $q_{ir}$. The stabilizers we have introduced here constitute the extreme case, namely the nodes $j$ for which $\tilde{\theta}_{rj} \equiv 0$. As we have shown, this 
requirement together with  the fact that $\tilde{\theta}_{rj}$ depends in turn on $q_{ir}$ has allowed us 
to extract the stabilization rules for crisply classified nodes, $q_{ir} = \delta_{g_i r}$. However, this concept could be relaxed to define stabilizers more generally by requiring only that $\tilde{\theta}_{rj} < \epsilon $ with 
$\epsilon$ appropriately chosen. 

It is possible to apply the notion of stabilizers to other probability models for node classification such as those considered by Airoldi {\it et al.}\cite{Airoldi} or Nowicki and Snijders~\cite{Snijders}. An inspection of 
Eqs.~(2) and (3) as well as the equation for $\hat{B}(g,h)$ 
of \cite{Airoldi} reveals a 
similar structure for the inter-relation between the edge-based class assignment probabilities $\phi$  and the class connection probability $\hat{B}(g,h)$, which are analogues of the probabilities $q$ and $\theta$. 
The variational Expectation Maximization approach of \cite{Airoldi} can also be applied to a 
 model that was considered by Nowicki and Snijders~\cite{Snijders}, which is more akin in spirit to 
the model presented here \cite{MM}. For both models, however the resulting rules of stabilization 
are rather involved due to the inclusion in the likelihood of the absence of edges, as well 
as due to the non-factorizable form of ${\rm Pr}(i \leftrightarrow j \vert g_i, g_j) \equiv \eta(g_i,g_j)$ as compared with Eq.~(\ref{eqn:thetacon}). The attractiveness of the probability model, Eqs.~(\ref{eqn:thetacon}) and (\ref{eqn:ProbA}), is that 
it delivers meaningful classifications despite of its simplicity, while at the same time 
the corresponding  stabilization rules have a rather immediate interpretation, as we have  
shown in Sections 3 and 4.

In summary, we have presented a general method for inferring information about which elements are most relevant in establishing group structures in a complex network. The maximum likelihood function upon which our inference is based is rather generic. This approach does not assume any additional {\it a priori} knowledge about the network, rendering it attractive in circumstances in which the available information about the nodes is limited. In particular, we have introduced the concept of stabilizers associated with a given NL-EM classification and shown that they play a central role in the network partition. If the stabilizers were removed from the network, the partition would lose its meaning. If on the other hand, the subgraph formed only by stabilizers is considered, the classification remain intact and useful information regarding the interaction between the different groups in the graph can be obtained. The stabilizers represent therefore the gist of a network partition. Their identification is highly useful in understanding the way in which the structure of complex systems form and their elements aggregate in clusters. This technique has a wide applicability as we have shown with three empirical examples of networks of very different origins: social sciences, semantics and ecology. In addition it raises several important questions, such as the role of these special nodes in the evolution of any dynamic process running on the graph such as the spreading of opinions, rumors or diseases, or even in the evolution of the graph itself if the network is dynamic. 

\ack We thank G. Bianconi, H. Bing\"ol, T. Cemgil, V. Colizza, A. Lancichinetti and F. Radicchi for useful comments. JJR and MM are funded by the project 233847-Dynanets of the EU Commission and Bo\u gazi\c ci University grant 08B302, respectively.

\appendix

\section{Numerical implementation details}

\subsection{NL-EM algorithm}
Given a network, the search for classifications of the NL-EM algorithm was 
carried out using an algorithm that alternates between simulated annealing
and a direct greedy iteration of Eqs.~(\ref{eqn:EM}) and (\ref{eqn:qir}). The program was run with a set of $10\, 000$ different initial conditions for $\theta$ and $\pi$ for each value of the number
of groups $\mathcal{N}_C$. Once the algorithm converged to a stationary value of the
likelihood function, the instance with the best $\bar{\mathcal{L}}$ was 
selected.

\subsection{Extraction of stabilizers}

We outline here the algorithm we have used to extract 
stabilizers from an NL-EM classification with strong nodes. 
The problem of determining the set of stabilizers associated with 
a strong node is related to the set covering problem in Computer Science, 
which is NP-complete. If a strong node has $s$ adjacent nodes with non-empty $\bar{\sigma}$, there are in principle $2^s$ 
combinations that have to be checked for finding the fundamental sets leading to stabilizations. In  
practice, many of the combinations can be eliminated by observing that if, say, $\bar{\sigma}_1,\bar{\sigma}_2, \ldots,  \bar{\sigma}_n$ have been selected as candidates for a stabilization, any $\bar{\sigma}$ that is a subset of the union of the $\bar{\sigma}_i$'s is redundant and thus cannot be part of that stabilization. This is the 
main strategy of our algorithm.  
Also note that, if there are $\mathcal{N}_C$ number of classes the number of possible distinct stabilizer sets $\bar{\sigma}$ is $\Sigma = 2^{\mathcal{N}_C - 1}$. For small $\mathcal{N}_C$,  $s$ can be larger 
than $\Sigma$ so that there are duplicates which can be removed beforehand.  

We have used a recursive algorithm for detecting the stabilizers. We partially order the $s$ sets $\bar{\sigma}_i$ by their size. Two binary arrays of size $s$, $iSelected$ and 
$iAvailable$ indicate the candidates already selected and those available 
for contribution to a stabilization, respectively. The classes against which the $s$ nodes 
stabilize are coded in an $s \times (\mathcal{N}_C-1)$ binary array $arrStab$, where the non-zero elements of 
$arrStab[j,*]$ indicate the classes against which node $j$ is stabilizing. 
A recursively called subroutine ${\bf\mbox{PickNext}}()$, givenin Fig.~A.1 in pseudo-code, performs the task of 
determining all stabilizations of a strong node, given the sets $arrStab$. 
In the algorithms we have assumed that there is already defined a procedure $\mbox{Where}(List,Value) = (Pointer, NFound)$, which takes a list and returns the indices where the list element equals to $Value$ along 
with the number of elements found $NFound$. Also in our notation when two lists are operated on term by term we denote this as $NewList[*] \gets ListOne[*] \;\; {\rm <Operator>} \;\; ListTwo[*]$, avoiding having 
to write out explicitly a loop over the operation on individual elements.

\begin{figure}[h!]
\begin{center}
\vspace{-0.1cm}
\includegraphics[width=\textwidth]{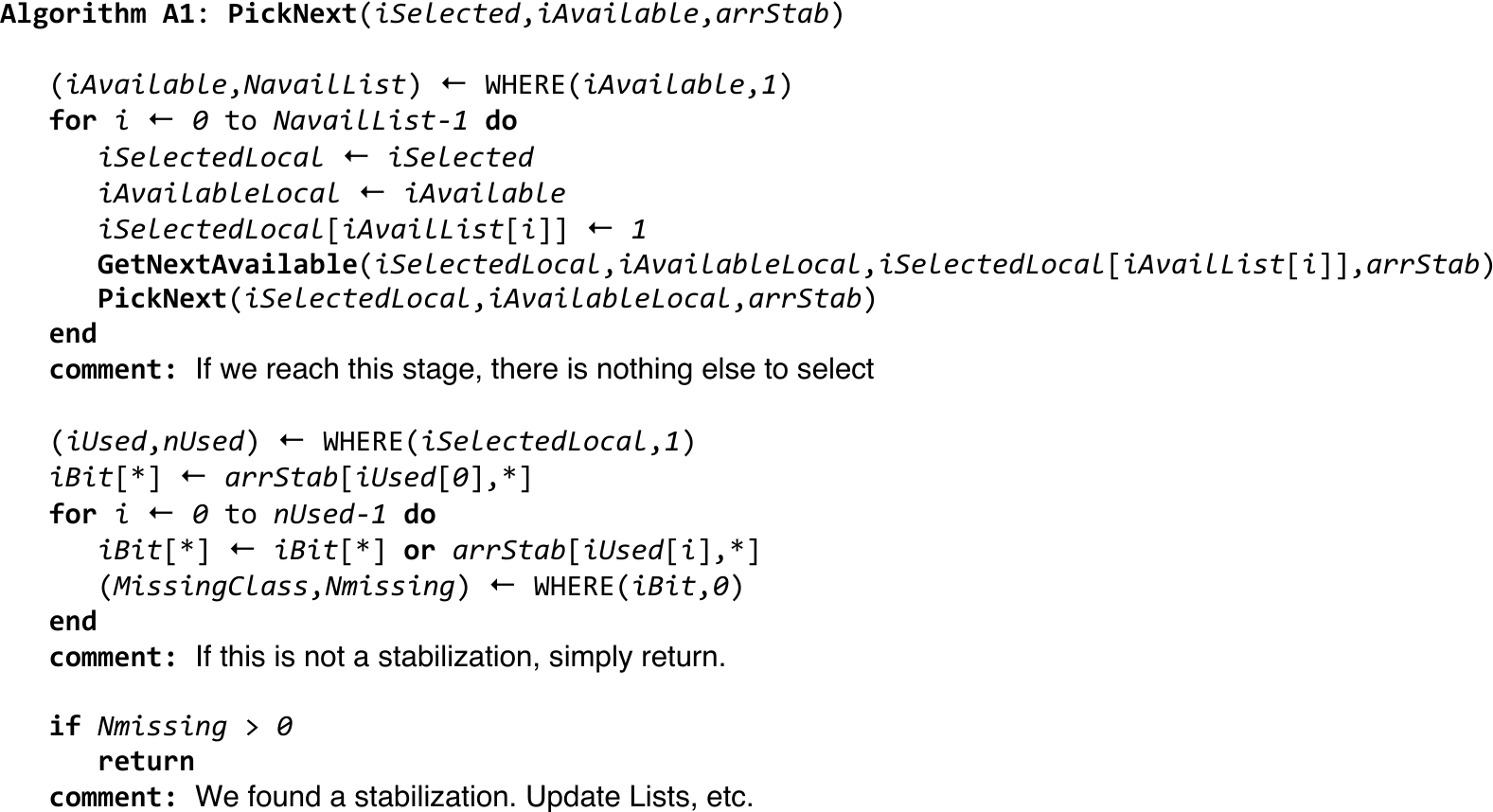}
\end{center}
\caption{Pseudocode for algorithm ${\bf\mbox{PickNext}}$.}
\label{fig:A1}
\end{figure}

Initially ${\bf\mbox{PickNext}}(iSelected,iAvailable,arrStab)$ is called with the binary arrays $iSelected$ and 
$iAvailable$ initialized to zero and one, respectively. The algorithm ${\bf\mbox{getNextAvailable}}{}$ (see Fig.~A.2) updates $iAvailableLocal$, the set of available stabilizers that can contribute to a stabilization after $iSelectedLocal[i]$ has been added.

\begin{figure}[h!]
\begin{center}
\vspace{-0.1cm}
\includegraphics[width=\textwidth]{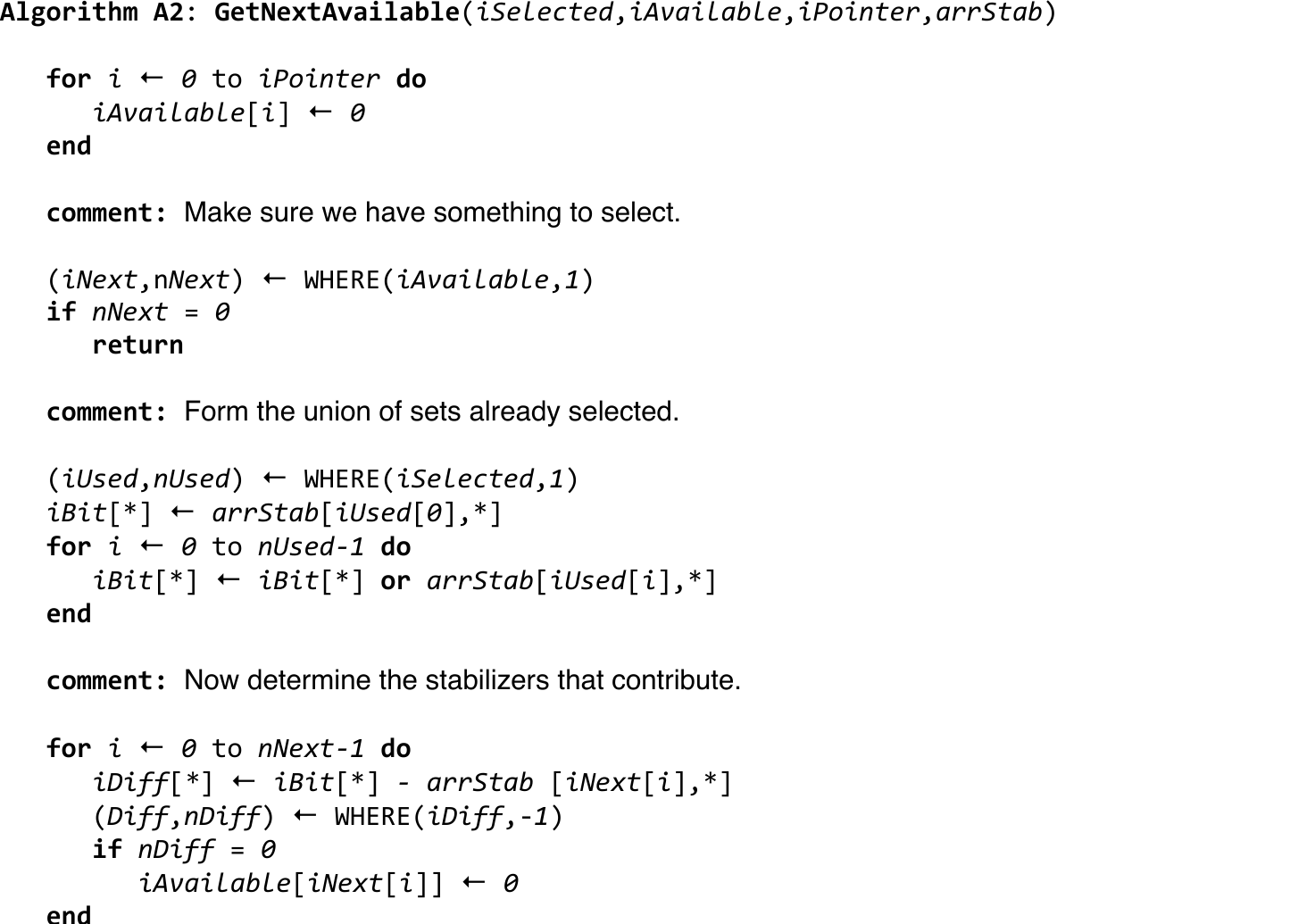}
\end{center}
\caption{Pseudocode for algorithm ${\bf\mbox{getNextAvailable}}$.}
\label{fig:A2}
\end{figure}

\section{Extension of NL-EM to directed graphs and stabilization}

A generalization of NL-EM to directed graphs that preserves structural equivalence~\cite{doreian,lorrain71,neo00} was recently 
provided in our earlier work \cite{us}. We assume that given a node $i$, a link to a node $j$ 
can be either out-going, in-going or bi-directional. We thus introduce the probabilities: 

\begin{itemize}
\item $\ra{\theta}_{rj}$ that a directed link leaving a vertex of group $r$ connects to node $j$, 
\item $\la{\theta}_{rj}$ that a directed link pointing to a node in group $r$ exists from $j$, and  
\item $\ta{\theta}_{rj}$ that a bidirectional link exiting from group $r$ connects to $j$, 
\end{itemize}
and construct the probability of realizing a directed graph $\mathcal{G}$  as 
\begin{equation}
{\rm Pr}(\mathcal{G},g|\pi,\la{\theta},\ra{\theta},\ta{\theta})  
= \prod_i \left[ \pi_{g_i} \prod_{ j \in \la{\nu}_i} 
\la{\theta}_{g_i,j} \, \prod_{ j \in \ra{\nu}_i} \ra{\theta}_{g_i,j} \,
\prod_{ j \in \ta{\nu}_i} \ta{\theta}_{g_i,j}  \right],  
\label{prob2} 
\end{equation}
$\la{\nu}_i$, $\ra{\nu}_i$, and $\ta{\nu}_i$ are the set of adjacent nodes of 
$i$ from which $i$ receives an in-coming, out-going, and bi-directional link, respectively. 
 
The likelihood can now be written as
\begin{equation}
\bar{\mathcal{L}}(\pi,\theta) = 
\sum_{ir} q_{ir} \left [ \ln{\pi_r} + 
\sum_{j \in \la{\nu}_i} \ln{\la{\theta}_{r,j}} + \sum_{j \in \ra{\nu}_i} \ln{\ra{\theta}_{r,j}} 
 + \sum_{j \in \ta{\nu}_i} \ln{\ta{\theta}_{r,j}} \right ], 
\label{lbardirected}
\end{equation}
which has to be maximized under the following constraint on the probabilities 
$\theta_{rj}$, 
\begin{equation}
\sum_i \left (\la{\theta}_{r,i}+ 
\ra{\theta}_{r,i} + \ta{\theta}_{r,i} \right ) = 1,
\label{eqn:thdirnorm}
\end{equation}
implying that there is no isolated node. The probability $\pi_r$, that a randomly 
selected node belongs to group $r$, is again given by $\sum_r \pi_r =1 $. 
The final result is~\cite{us}
\begin{equation}
\pi_r = \frac{1}{N} \sum_i q_{ir},
\end{equation} 
\begin{eqnarray}
\la{\theta}_{rj} = \frac{\sum_{i \in \ra{\nu_j}} q_{ir} }{\sum_i q_{ir} ( \bar{k}_i^i + \bar{k}_i^o - 
\bar{k}_i^b )}, \nonumber \\
\ra{\theta}_{rj} = \frac{\sum_{i \in \la{\nu_j}} q_{ir}}{\sum_i q_{ir} ( \bar{k}_i^i + \bar{k}_i^o - 
\bar{k}_i^b )}, \label{eqn:thetarj} \\ 
\ta{\theta}_{rj} = \frac{\sum_{i \in \ta{\nu_j}} q_{ir} }{\sum_i q_{ir} ( \bar{k}_i^i + \bar{k}_i^o - 
\bar{k}_i^b )} \nonumber  ,
\end{eqnarray}
where $\bar{k}_i^i$, $\bar{k}_i^o$ and $\bar{k}_i^b$ are the in-degree,
out-degree and bi-directional degree of node $i$, respectively. 

These expressions have to be again supplemented with the self-consistent equation for
$q_{ir}$ which now reads 
\begin{equation}
q_{ir} = \frac{ \pi_r 
\prod_{j \in \la{\nu_i}} \la{\theta}_{rj} \,
\prod_{j \in \ra{\nu_i}} \ra{\theta}_{rj} \,
\prod_{j \in \ta{\nu_i}} \ta{\theta}_{rj} \,
}
{
\sum_s \left \{ \pi_s 
\prod_{j \in \la{\nu_i}} \la{\theta}_{sj} \,
\prod_{j \in \ra{\nu_i}} \ra{\theta}_{sj} \,
\prod_{j \in \ta{\nu_i}} \ta{\theta}_{sj} \right \}
}.
\label{eqn:q_ir}
\end{equation}

Note that when we have only bi-directional links so that $\la{\nu}_i = \ra{\nu}_i = \emptyset$ for 
all $i$, and it follows from Eq.~(\ref{eqn:thetarj}) that 
$\la{\theta}_{rj} = \ra{\theta}_{rj} = 0$. Thus we recover the undirected EM equations
Eqs.~(\ref{eqn:EM}) and (\ref{eqn:qir}) under the identification 
$\theta_{rj} = \ta{\theta}_{rj}$.

\subsection{Stabilization rules for directed graphs}

The case of directed graphs is similar to the undirected case with a few 
minor modifications. Given a NL-EM classification of a directed graph $\mathcal{G}$, 
we associate with each node $i$ the following four sets:
\begin{itemize}
\item $\overline{\la{\sigma}}_i = \{r \vert \la{\theta}_{ri} = 0\}$, the set of groups that $i$ does not 
have an out-going connection to,
\item $\overline{\ra{\sigma}}_i = \{r \vert \ra{\theta}_{ri} = 0\}$, the set of groups that $i$ does not 
have an in-going connection to,
\item $\overline{\ta{\sigma}}_i = \{r \vert \ta{\theta}_{ri} = 0\}$, the set of groups that $i$ does not 
have an bi-directional connection to,
\item $\overline{c}_i = \{r \vert q_{ir} = 0\}$, the set of groups that $i$ does not belong to,
\end{itemize}
along with their complements, $\la{\sigma}_i$, $\ra{\sigma}_i$, $\ta{\sigma}_i$, and $c_i$.

The NL-EM equations, Eqs.~\ref{eqn:thetarj} and \ref{eqn:q_ir}, relate  
the sets $\overline{\sigma}_i$ and $\overline{c}_i$ to each other as follows:
\begin{eqnarray}
\bigcup_{j \in \la{\nu}_i} \; \overline{\la{\sigma}}_j \;
\bigcup_{j \in \ra{\nu}_i} \; \overline{\ra{\sigma}}_j \;
\bigcup_{j \in \ta{\nu}_i} \; \overline{\ta{\sigma}}_j \;
 &=& \overline{c}_i, \label{eqn:dirstab1}\\
\bigcap_{i \in \ra{\nu}_j} \; \overline{c}_i &=& \overline{\la{\sigma}}_j, \label{eqn:dirstab2}\\
\bigcap_{i \in \la{\nu}_j} \; \overline{c}_i &=& \overline{\ra{\sigma}}_j, \label{eqn:dirstab3}\\
\bigcap_{i \in \ta{\nu}_j} \; \overline{c}_i &=& \overline{\ta{\sigma}}_j. \label{eqn:dirstab4}
\end{eqnarray}

Defining the set of all stabilizer classes associated with a node, irrespective of the directionality as 
\begin{equation}
\overline{\sigma}_i \equiv  \overline{\la{\sigma}}_i \cup \overline{\ra{\sigma}}_i \cup
\overline{\ta{\sigma}}_i,  
\end{equation}
the stabilization condition for a node $i$ becomes identical to the one for the undirected case,  
\begin{equation}
  \bigcup_{j \in \nu_i } \; \bar{\sigma}_j = \bar{c}_i.
\end{equation}


\begin{thebibliography}{99}
\bibitem{NOTE}
The question of whether a given graph admits a NL-EM solution 
that is a (strong) partition of all of its nodes into $\mathcal{C}$ groups, can thus be seen as 
a coloring problem: We seek a partition $g$ of the nodes, i.e. a coloring $g_i$, corresponding to $c_i = \{g_i\}$, such that 
\begin{equation*}
 \bigcup_{j \in \nu_i}  \bigcap_{\ell \in \nu_j} \; \bar{c}_\ell = \bar{c}_i,
\end{equation*}
is satisfied for all $i$.
\end{thebibliography}
\end{document}